\documentclass[prl,amsmath,amssymb,twocolumn]{revtex4}

\usepackage{graphicx}% Include figure files
\usepackage{dcolumn}% Align table columns on decimal point
\usepackage{bm}% bold math
\usepackage{epstopdf}
\usepackage{epsfig}
\usepackage{multirow}

\usepackage{color}
\usepackage{ulem}

\begin{document}

\title{An Exchange-Coupled Donor Molecule in Silicon}

\author{M. Fernando Gonz\'alez-Zalba}
\email{mg507@cam.ac.uk}
\affiliation{Hitachi Cambridge Laboratory, Cavendish Laboratory, UK}
\affiliation{Cavendish Laboratory, University of Cambridge, Cambridge, UK}
\author{Andr\'e Saraiva}
\email{also@if.ufrj.br}
\affiliation{Instituto de Fisica, Universidade Federal do Rio de Janeiro, Caixa Postal 68528, 21941-972 Rio de Janeiro, Brazil}
\author{Mar\'\i a J. Calder\'on}
\affiliation{Instituto de Ciencia de Materiales de Madrid, ICMM-CSIC, Cantoblanco, E-28049 Madrid, Spain}
\author{Dominik Heiss}
\affiliation{Cavendish Laboratory, University of Cambridge, Cambridge, UK}
\affiliation{Present address: Eindhoven University of Technology, The Netherlands}
\author{Belita Koiller}
\affiliation{Instituto de Fisica, Universidade Federal do Rio de Janeiro, Caixa Postal 68528, 21941-972 Rio de Janeiro, Brazil}
\author{Andrew J. Ferguson}
\affiliation{Cavendish Laboratory, University of Cambridge, Cambridge, UK}
\date{\today}

\maketitle

{\bf Donors in silicon, conceptually described as hydrogen atom analogues in a semiconductor environment, have become a key ingredient of many \lq\lq More-than-Moore\rq\rq\ proposals such as quantum information processing~\cite{kane_silicon_1998,Hollenberg2004,morello2010,pla2012,pla2013} and single-dopant electronics~\cite{Koenraad,zwanenburg_RMP2013}. The level of maturity this field has reached has enabled the fabrication and demonstration of transistors that base their functionality on a single impurity atom~\cite{lansbergen-NatPhys,fuechsle2012} allowing the predicted single-donor energy spectrum to be checked by an electrical transport measurement. Generalizing the concept, a donor pair may behave as a hydrogen molecule analogue. However, the molecular quantum mechanical solution only takes us so far and a detailed understanding of the electronic structure of these molecular systems is a challenge to be overcome. Here we present a combined experimental-theoretical demonstration of the energy spectrum of a strongly interacting donor pair in the channel of a silicon nanotransistor and show the first observation of measurable two-donor exchange coupling.
Moreover, the analysis of the three charge states of the pair shows evidence of a simultaneous enhancement of the binding and charging energies with respect to the single donor spectrum. %Such enhancement suggests a novel physical mechanism to increase the operation temperature of conventional single-atom transistors and improve their robustness against interfacial electric fields. 
The measured data are accurately matched by results obtained in an effective mass theory incorporating the Bloch states multiplicity in Si, a central cell corrected donor potential and a full configuration interaction treatment of the 2-electron spectrum. 
%The bond length of the donor "molecule" is identified with an interatomic separation of $2.3 \pm 0.5$ nm.  
Our data describe the basic 2-qubit entanglement element in Kane's quantum processing proposal~\cite{kane_silicon_1998}, namely exchange coupling, implemented here in the range of molecular hybridization.}

%D$_2^{\{0, +, 2+\} }$
%D$^{\{-,0,+\}}$

%{\bf Substitutional donors in Si are analogous to hydrogen atoms with giant (40 times larger)
%electronic orbitals~\cite{Kohn1955a}. Its extent is enough to integrate a single impurity to electronic circuitry~\cite{zwanenburg_RMP2013}.
%Donor molecules are a natural continuation of this concept, with potential applications
%as components of quantum devices~\cite{kane_silicon_1998,Hollenberg2004}. But the analogy with traditional quantum
%chemistry only takes us so far, and a detailed understanding of the
%electronic structure of these molecular systems is a challenge yet to be
%overcome. By measuring the spectrum of a low-doped sample we identify here an isolated
%strongly interacting donor pair, with measurable exchange coupling. The pair geometry is inferred by comparing the measured spectrum and an effective mass theory
%incorporating the Bloch states multiplicity in Si, a central cell corrected
%donor potential and full configuration interaction. From our results we discuss
%the conditions for optimal performance of donor based quantum electronics.}

A single dopant in silicon has the ability to dramatically alter the electrical properties of state-of-the-art CMOS transistors \cite{Asenov2003, Shinada2005, pierre2009} opening up a window for fundamental innovation. Single dopants have been detected via resonant transport in the sub-threshold regime of nanoFETs \cite{Sellier2006, Ono2007, Tan2009} and recently a single-atom transistor has been fabricated deterministically \cite{fuechsle2012}. Experimental studies have been backed up by theoretical calculations that explained deviations from the donor bulk energy spectrum.  Capacitive coupling to the gate electrodes \cite{Sellier2006}, electric-field-induced Stark shift \cite{Rahman2011a} and dielectric confinement \cite{calderon-Dminus-2010} modify the one and two-electron binding energies and reduce the charging energy presenting a challenge for future technologies in terms of reproducibility and elevated temperature operation.

Two-donor devices present an opportunity to harness the potential of single donor technology. In that sense, researchers have developed donor-based single-electron pumps~\cite{Roche2013,Lansbergen2012,Tabe2010} and studied fundamental properties of donors such as valley-orbit splitting \cite{Roche2012}, Anderson-Mott transition \cite{Prati2012a} and coherent coupling \cite{Dupont2013}. However, the most appealing implementation is quantum computation where, in the Kane model, a donor molecule forms the basic unit of quantum information processing \cite{kane_silicon_1998}.

In this Letter we present transport spectroscopy through an arsenic molecule in silicon and develop a novel analogy between donor dimers and the hydrogen molecule based on a central-cell-corrected effective-mass approximation. The model accounts for the enhanced binding and charging energies, the finite exchange coupling and the robustness against electric-field detuning. Previous attempts based on EMA studies failed to capture the subtle effects of valley physics in silicon~\cite{Miller1960} and only the asymptotic behaviour at large inter-donor distances was obtained through Heitler-London method~\cite{koiller_exchange_2001}. The size of the system hinders first principles and tight binding studies due to computational cost~\cite{zhang2013,Rahman2011}

We find that our measured spectrum is in excellent agreement with the theoretical model for an As molecule with inter-atomic distance R=2.3 $\pm$0.5~nm, demonstrating the first experimental evidence of transport through a donor molecule in silicon.

\begin{figure}[h] %Fig. 1
%\vspace{50 mm}
\includegraphics[scale=0.5]{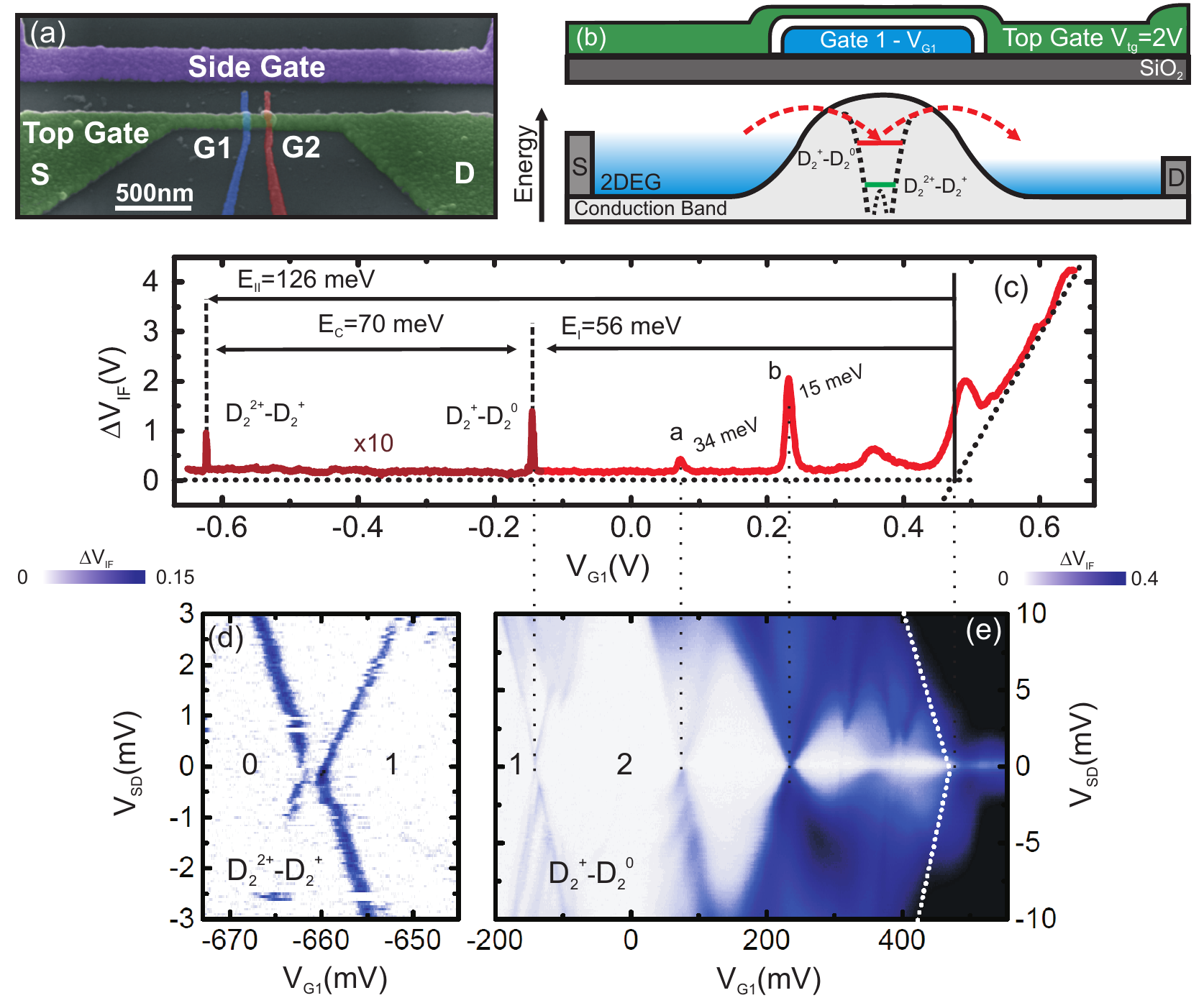}
\centering
\caption{\textbf{Device structure and spectroscopy.} (a) False colour scanning electron microscope image of the device. (b) Schematic cross-section of the energy bands across the relevant tunnel barrier G1 along the direction of transport indicating the potential induced by the molecule. (c) Reflectrometry response V$_{IF}$ as a function of the source barrier gate in the sub-threshold regime.  (d) The signal of the zero to one electron transition, D$_2^{2+}\rightarrow$D$_2^{+}$, and (e) of the one to two electron transition, D$_2^{+}\rightarrow$D$_2^{0}$, with respect to the conduction band edge, white dotted line.}
\label{fig1}
%\vspace{50 mm}
\end{figure}

In this study we use a double-gated metal-oxide-semiconductor field-effect transistor to map the energy spectrum of a donor molecule in Si~\cite{gonzalez-zalba-NJP2012, Angus2008}, as seen in Fig.~\ref{fig1}(a). The nanoFETs G1 and G2 control the energy band bending of the environment immediately under it. Setting it below threshold generates a barrier and current only flows through quantum tunneling (Fig.~\ref{fig1}(b)). Furthermore, G1(2) tunes the electrochemical levels of the impurities, \textit{i.e},  the energy required to add an extra electron to the system. Whenever one of these levels resides within the energy window of the bias voltage (centered at V$_{SD}$), transport through the structure can occur and resonant tunneling current peaks appear. In order to reduce the effect of $1/f$ noise on these devices we used radio-frequency reflectometry~\cite{Schoelkopf1998}. This technique probes the complex impedance of the device generating a DC output voltage $V_{IF}$ which is proportional to the differential conductance (see Methods).

Up to 12 devices measured at 4~K showed subthreshold resonances associated to individual As atoms --- charging energies $E_{C}$ between 23 and 37~meV, as previously reported for gated donors~\cite{calderon-Dminus-2010, Tan2009} --- with an average of 2 As atoms per transistor. The FET controlled by G1 was the only one presenting resonant lines with enhanced charging and binding energies, signaling a strongly coupled donor pair in the channel of the transistor.

Fig.~\ref{fig1}(c) shows the rf-response as a function of the gate voltage ($V_{G1}$). Below threshold ($V_{G1}$=470~mV), obtained from a fit to the linear region of the FET, we observe a set of single donor or unintentional quantum dot lines, marked as a and b, located at 34 and 15 meV with respect to the conduction edge (see Methods). Notably, we observe two other transitions, D$_2$$^{2+}$$\rightarrow$D$_2$$^+$ and D$_2$$^{+}$$\rightarrow$D$_2$$^0$ at -660~mV and -140 mV respectively, which are the main focus of this paper.

We quantify these energies by measuring the characteristic Coulomb diamonds (Figs.~\ref{fig1}(d) and~\ref{fig1}(e)). We find a charging energy of $E_C$=70$\pm$3~meV, read from the $V_{SD}$ axis at the point where the edges of the diamonds meet as to avoid the use of gate-voltage-to-energy conversion parameters that depend on electron occupancy and bias conditions (see Methods). Moreover, the first ionization energy $E_{\rm I}$=56$\pm$5~meV is obtained from the point at which the D$_2^+$$\rightarrow$D$_2^0$ transition meets with the edge of the conduction band, indicated by the dotted white lines. The second ionization energy is therefore $E_{\rm II}$=126$\pm$6~meV. These values are markedly larger than expected for an isolated arsenic donor in bulk silicon:  $E_C$=51.71~meV, $E_{I}$=2.05~meV (As$^0$$\rightarrow$As$^-$), and $E_{II}$=53.76~meV (As$^+$$\rightarrow$As$^0$)~\cite{calderon-Dminus-2010}.

\begin{figure} %[h]
%\vspace{30 mm}
\includegraphics[scale=0.16]{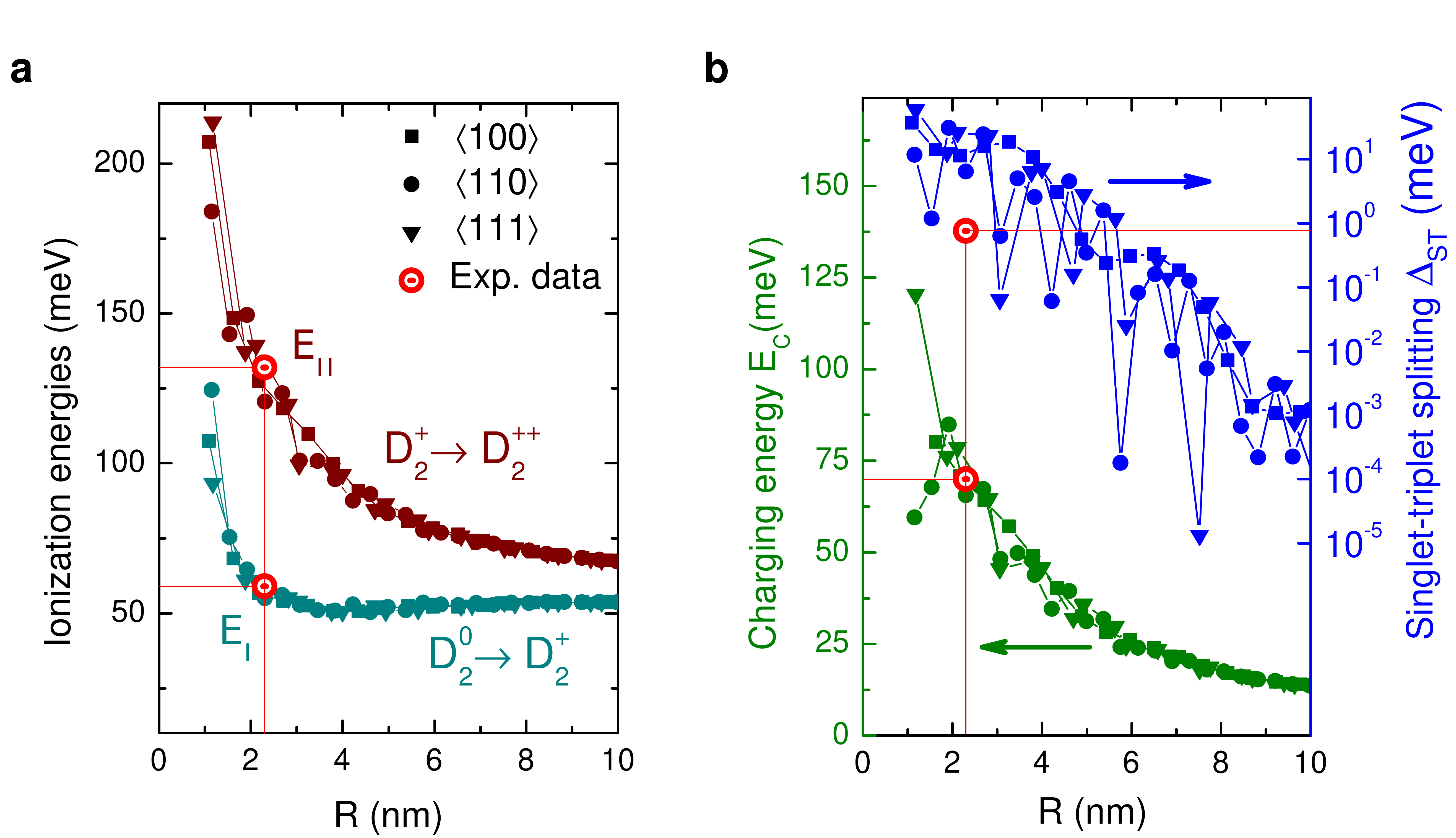}
\centering
\caption{\textbf{Analysis of interdonor distance.} Calculated dependence on the interdonor distance $R$ of (a) the first ionization energy and second ionization energy, (b) the charging energy (left) and the singlet-triplet splitting (right). The experimentally measured energies are shown (red circles) at the estimated interdonor distance $R=2.3\pm 0.5$ nm.}
\centering
\label{fig2}
%\vspace{50 mm}
\end{figure}

The naive comparison of the measured values to a rescaled theory of H$_2$ molecules also leads to disappointing results (see Methods). Overcoming these apparent inconsistencies demands a realistic theory of donor pairs in Si. Calculations are performed here within an improved effective mass approach, which includes central cell corrections with an empirical radius $r_c$ leading to the correct A$_1$ energies, the full valley structure of the Si conduction band, and electron-electron correlations through a configuration interaction method. This approach extends previous results~\cite{kane_silicon_1998,koiller_exchange_2001} to much smaller distances $R$ between dopants with higher accuracy. The central cell correction enhances the one and two electron binding energies (Fig.~\ref{fig2}(a)) and indirectly enhances the electron-electron repulsion shown in Fig.~\ref{fig2}(b), due to the tighter confinement of the A$_1$ states.

Comparison between theory and experiment leads to the identification of the particular interdonor distance $R$ of the molecule. We first compare the ionization energies because of the smooth behavior of these energies with $R$, regardless of the molecular orientation. Both $E_I$ and $E_{II}$ measurements, indicated in Fig.~\ref{fig2}(a), are consistent with an inter-donor distance of $R=2.3\pm 0.5$ nm. This result is also in agreement with the charging energy profile shown in Fig.~\ref{fig2}(b) despite its oscillatory behaviour. 

\begin{figure}[h]
%\vspace{30 mm}
\includegraphics[scale=0.5]{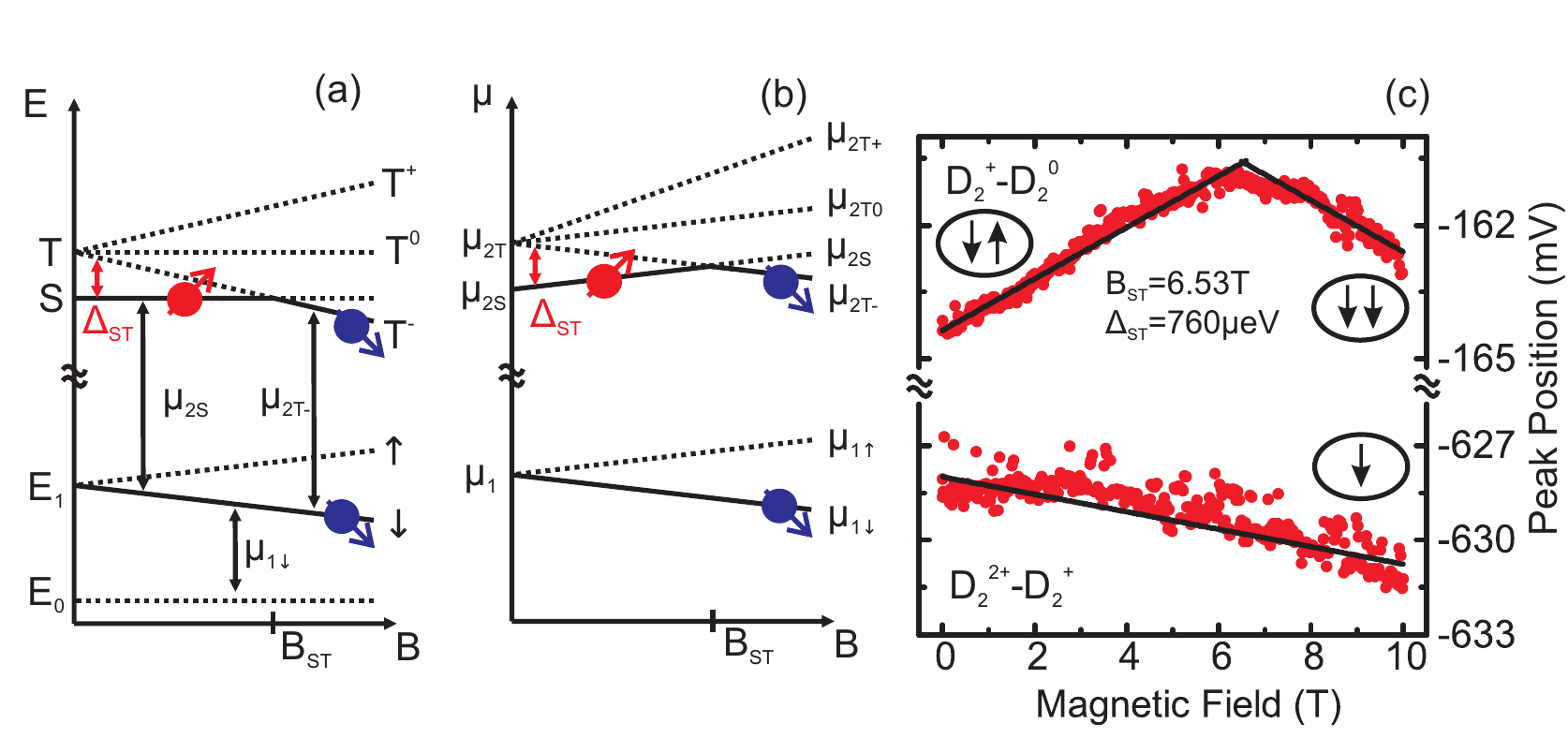}
\centering
\caption{\textbf{Spin filling and exchange coupling.} Schematic representation of (a) the energy evolution with field for the one-electron and the two-electron states and (b) the evolution of the electrochemical levels $\mu_{1(2)}$ as a function of magnetic field.  $\Delta_{ST}$ is the singlet-triplet splitting at zero field. (c) Measured magnetic field dependence of the D$_2^{2+}\rightarrow$D$_2^{+}$ transition (bottom) and the D$_2^{+}\rightarrow$D$_2^{0}$ transition (top) showing a singlet-triplet crossover. Black solid lines are guides to $(\pm)g\mu_B/2$. }
\label{fig3}
%\vspace{50 mm}
\end{figure}

We may confirm the molecular nature of these states and learn about the spin configuration and the singlet-triplet splitting $\Delta_{ST}$ from the spin filling sequence of the one and two-electron energy states [Fig.~\ref{fig3}(a)]. The exchange coupling $J$ in the Heisenberg spin Hamiltonian $J \mathbf{S_1} \cdot \mathbf{S_2}$ -- the main ingredient in CNOT operations for Kane's qubits --  is determined by the singlet-triplet splitting as $J = \Delta_{ST}/\hbar^2$. Experimentally,  $\Delta_{ST}$ can be accurately identified by monitoring the evolution of the electrochemical potentials ($\mu_{1,2}$) as a function of magnetic field, in this case applied in the plane of the device (Fig.~\ref{fig3}(b)). The rate of change with field is given by \cite{Lim2011},
\begin{equation}
	\frac{\partial \mu_{1,2}}{\partial B_\parallel}= -g\mu_B\Delta S_{1,2}
\end{equation}
where $g$ is the g-factor~$\approx$~2 for donor-bound electrons in Si, $\mu_B=$ 57.8~$\mu$eV/T the Bohr magneton and $\Delta S$ the change in total spin of the molecule when an extra electron is added. Filling the molecule with a spin-down (up) electron results in a $-(+)g \mu_B/2$ slope of the chemical potential with magnetic field. In Fig.~\ref{fig3}(c) we measure the evolution of $\mu_{1(2)}$ as a function of magnetic field up to 10~T.

The ground state of the first electron is the lower Zeeman branch, therefore the 0 to 1 electron transition shifts down as a function of field. In Fig.~\ref{fig3}(c), the data is compared to a Zeeman shift $\mu_B$, calibrated to voltage shift using the lever arm $\alpha _{0\rightarrow 1}=0.21$ extracted from the slope of the edges of the D$_2$$^{2+}\rightarrow$D$_2$$^+$ transition, confirming the loading of a spin-down electron.

Experimentally the two-electron ground state Fig.~\ref{fig3}(c) shifts in the opposite direction at low magnetic fields, consistently with the loading of a spin-up electron and a singlet two-electron ground state at low fields. Here, we use $\alpha _{1\rightarrow 2}=0.10$ extracted from the slopes of the D$_2$$^{+}\rightarrow$D$_2$$^0$ transition. At $B_{ST}$=6.53~T the slope changes to $-g\mu_B/2$ indicating a change in spin configuration, a transition from spin-singlet to spin-triplet two electron ground state. From the magnetic field at which this happens, we extrapolate a zero field singlet-triplet splitting $\Delta_{ST}$ of 0.76~meV.

The small singlet-triplet splitting $\Delta_{ST}$ measured can be explained from the particular molecule orientation leading to destructive interference of the electronic wavefunctions as seen in Fig.~\ref{fig2}(b) for the $\left\langle 110 \right\rangle$ and $\left\langle 111 \right\rangle$ crystallographic directions. This is a vivid example of the fragility of the singlet-triplet splitting to the particular positioning of donors, as predicted in Ref.~\cite{koiller_exchange_2001}.

\begin{figure}[h]
\includegraphics[scale=0.18]{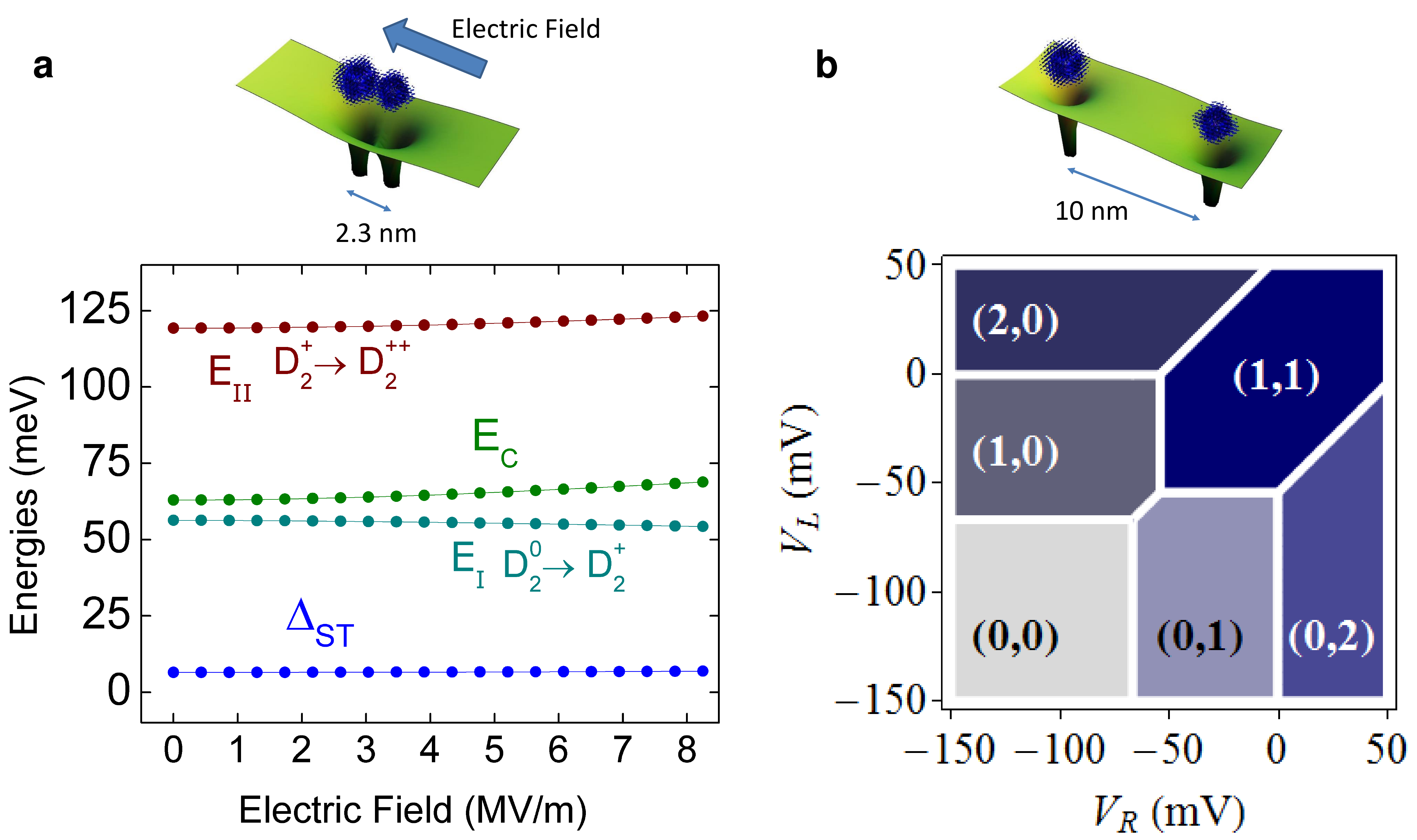}
\centering
\caption{\textbf{Electric Field Effects.} (a) Impact of an electric field on the spectrum of a donor pair at $R=2.3$ nm (similar to our sample). Up to fields as high as the SiO$_2$ dielectric strength, the charging energies, first and second ionization energies and the singlet-triplet splitting remain essentially unchanged. (b) For $R=10$nm, the charge distribution is quite sensitive to external fields, as implied by the charge diagram obtained. The pairs is aligned along the (110) direction in both calculations.}
\label{fig4}
\end{figure}

Finally we investigate the role of external electric fields since these are known to modify the single-donor spectrum \cite{Rahman2011a}. Figure~\ref{fig4}(a) shows the effect of a detuning electric field $E=(V_R-V_L)/R$ applied along the axis of the molecule for the distance $R=2.3$ nm. This is accomplished by shifting the on-site energies on the molecular orbital theory by $\delta_{L,R}= - e V_{L,R}$. We disregard here the proximity of interfaces~\cite{calderon-Dminus-2010, Tan2009}.
The separation between donors is too small for the electric field to generate a significant detuning. Moreover, the ionization energies are large due to the molecular hybridization, so that the charge states are robust against these external fields. This robustness explains why our theory for bulk Si describes well the energy levels even under the complex environment in our FET devices. %This is an appealing feature for the scalable fabrication of donor-based transistors.

We also study the hypothetical donor-pair separation, $R=10$ nm. In this case, the existence of a classically forbidden region between the two donor sites permits a study of the charge occupation of each site separately. The charge stability diagram, shown in Fig.~\ref{fig4}(b), is consistent with the data from Fig.~\ref{fig2} and equivalent to the diagram in a double quantum dot~\cite{hanson2007}. The analogy with quantum dots may be explored to implement charge and spin qubits.

The control over the charge degree of freedom at the $(0,1) \leftrightarrow (1,0)$ transition, instrumental for charge qubit proposals~\cite{Hollenberg2004}, could be implemented with modest electric fields. We estimate the tunnel coupling to be $t =210~\mu$eV from the level anti-crossing around $\delta_{L}=\delta_{R}=60$~meV.

Interestingly, it is possible to estimate the order of magnitude of the singlet-triplet splitting using the Hubbard model in the limit of small tunneling and large charging energy. This regime (Mott limit) leads to $\Delta_{ST}\approx t^2 /U = 0.8~\mu$eV, estimating the same-site charging energy $U=52$~meV from the bulk D$^-$ energy in As. The actual value calculated within the full configuration interaction method is $\Delta_{ST}=1.2 \mu$eV. We expect this approximation to be valid only for $R\gtrsim 8$~nm where a classically forbidden barrier between the sites exists.

%The (2,0) and (0,2) configurations are slightly more bound than a bulk D$^-$ state~\cite{calderon-Dminus-2010}. The presence of a positive charge (empty donor site) in the vicinity of the doubly occupied donor stabilizes the single electron states, but leads to a more compact electronic cloud, so that the charging energy is enhanced and the second electron is not as stable as a single particle calculation would suggest. This makes clear that it is important to explicitly calculate the effects of the electron-electron repulsion and the Pauli exclusion.

Our measurements and theory offer a proof of concept for quantum computation proposals relying on the exchange coupling between electron spins on donors. By analogy with artificial molecules, e.g. GaAs and Si double quantum dots, there is potential for a dopant molecule to be operated as a singlet-triplet spin qubit overcoming current difficulties with Kane's architecture exchange gates. Such a system would have the dual advantages of enabling spin manipulation via electric, rather than magnetic fields, and the long spin coherence times of dopants in silicon. Moreover, the sample we measure here has an electron-electron spin exchange coupling $J$ much larger than the hyperfine coupling $A$ with the nuclei, akin to real molecules. This raises the possibility of engineering longer and more complex chains as a playground for \textit{quantum chemistry on a chip}. The exchange coupling becomes comparable to the As bulk hyperfine coupling $A=198$ MHz at an interdonor distance of $R\approx 10$~nm, which sets the ideal geometry for Kane's architecture.

\section{Methods Summary}
{\bf Experiment.} Devices were fabricated on a high-resistive silicon wafer ($\rho$$>$7000~$\Omega$cm) and low-dose-ion-implanted with arsenic (D=1e11~cm$^{-2}$, $E$=15~keV). The sacrificial oxide was removed after implant and a 10~nm SiO$_2$ gate oxide was regrown at 850$^\circ$C. C-V measurements of similar devices indicate an interface trap charge density of 1.8$\times$10$^{10}$~cm$^{-2}$. The calculated profile gives a dopant density of 5$\times$$10^{16}$cm$^{-3}$ at 10~nm from the interface. Two sets of electrically independent e-beam-defined aluminium gates form the structure with nominal channel dimensions l=40~nm and w=120~nm. A positive voltage on the top gate induces an electron accumulation layer which constitutes the source and drain reservoirs of the FET. Measurements were performed at the base temperature of a dilution refrigerator with an electron temperature of 200~mK. Radio-frequency reflectometry was performed at 360~MHz by embeding the sample in a rf-tank circuit formed by a surface mount inductor (390~nH) and the parasitic capacitance to ground (500~fF).

{\bf Theory.} We adopt the multivalley effective mass theory with the explicit inclusion of a Yukawa-type potential with empirical cut-off radius accounting for the central cell corrections. The effective mass is taken to be isotropic and  to reproduce the hydrogenic energy obtained from the anisotropic Kohn-Luttinger model. The envelope function is obtained separately for each of the A$_1$, T$_2$ and E states variationally as 1s Slater-type orbitals. All single-electron integrals are calculated analytically following Ref.~\cite{Tai1986}. Electron-electron repulsion integrals are calculated within Pople's STO-2G scheme~\cite{Hehre1969}. The wavefunction and energy spectrum of the two-electron problem is obtained through spin-adapted configuration interaction (SACI), an exact theory within the chosen basis set which permits the separate analysis of the singlet and triplet subspaces. The two-electron wavefunction radii are obtained minimizing separately the singlet and triplet lowest eigenvalues of the SACI matrices.

\section{Acknowledgements}

The authors thank P.A. Gonzalez for fruitful discussions. MFGZ and AJF acknowledge support from EPSRC grant no. EP/H016872/1 and the Hitachi Cambridge Laboratory. AJF was supported by a Hitachi Research Fellowship. MFGZ was supported by an Obra Social La Caixa fellowship and Gobierno de Navarra Research grant. AS and BK  performed this work as part of the Brazilian National Institute for Science and Technology on Quantum Information and also acknowledge  partial support from the Brazilian agencies FAPERJ, CNPq, CAPES. MJC acknowledges support from MINECO-Spain through Grant FIS2012-33521.

\section{Author contributions}
MFGZ and AS contributed equally to this work.
MFGZ, DH and AJF fabricated the device, performed the measurements and interpreted the data. AS, MJC and BK devised the model. AS and MJC performed the calculations. MFGZ and AS produced the figures. All authors participated in the writing of the paper.

\section{Additional information}

The authors declare no competing financial interests.

\section{Methods}
{\bf Device fabrication.} The device fabrication starts by growth of a 10~nm sacrificial oxide on a high-resistivity ($\rho$$>$7000~$\Omega$cm) (100) silicon wafer. Ohmic contacts were defined by optical lithography and ion implantation of phosphorus (15~keV, $10^{15}$~cm$^{-2}$) and dopants included by low-dose (15~keV, $10^{11}$~cm$^{-2}$) ion-implantation of arsenic.  The sacrificial silicon oxide was removed after the implant and a 10~nm SiO$_2$ gate oxide was regrown at 850$^\circ$C, which also anneals out the implantation damage. A forming gas anneal at 450$^\circ$C for 30~min was performed followed by a rapid thermal anneal for 15~s at 1050$^\circ$C to reduce the interface trap and fixed oxide charge density.  The As profile was calculated by an implantation Monte-Carlo simulator \cite{SRIM} and has a maximum at 10~nm from the interface and a peak density of 5$\times$10$^{16}$cm$^{-3}$ resulting in a statistical mean inter-dopant distance of 27~nm. However, the expected segregation of As towards the interface during the subsequent thermal processing leads to an increase in the donor density at the interface and favours clustering~\cite{Grove1964}. By comparison, the residual phosphorus doping is estimated to be smaller than 10$^{12}$~cm$^{-2}$.

Subsequent to the silicon processing, surface gates were fabricated by electron beam lithography and thermal evaporation of aluminium. Fig.\ref{fig1}(a,b) shows a scanning electron microscopy image of an identical device and its schematic cross-section. As a first step, two gates 40~nm wide and 100~nm apart were defined by evaporation of a 25~nm thick layer of aluminium.  They are used to form the tunable source and drain tunnel barriers. After thermal oxidation at 150$^\circ$C for 5~min creating a 5~nm AlO$_x$ layer, a second electrically-independent 60~nm thick top gate was deposited over the barriers. This top-gate defines the 120~nm wide channel of the SET and leads that overlap with the doped ohmic contacts. From the lithographical dimensions of the two sets of gates it is possible to estimate an average number of 2 As atoms per tunnel barrier. Prior to measurement the samples undergo a nitrogen ambient post-fabrication anneal at 350$^\circ$C for 15~min. The interface trap density, measured on simultaneously processed field effect transistors, without e-beam radiation, by means of low frequency split C-V method \cite{Koomen1973}, is 1.8$\times$10$^{10}$cm$^{-2}$.

\begin{figure}[h]
	\centering
		\includegraphics[scale=0.5]{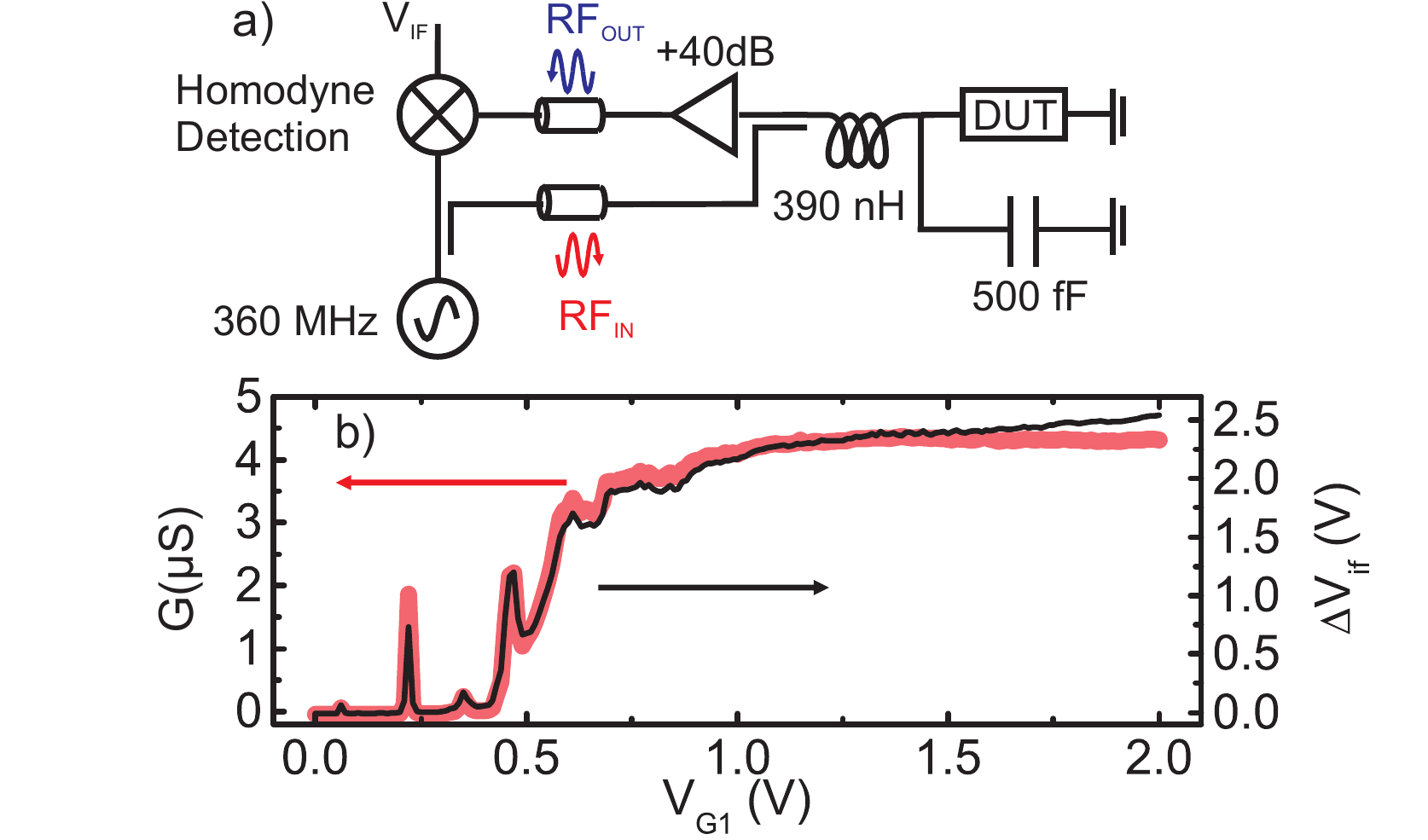}
	\caption{\textbf{Radio-frequency reflectometry.} (a) RF setup. The source of the sample is connected to a resonant tank circuit form by surface mount inductor (390~nH) and the parasitic capacitance to ground (500~fF). +40~dB amplification at 1.5~K is achieved by a cryogenic amplifier. (b) Comparison between the reflectometry response $\Delta V_{IF}$ and the source-drain conductance as a function of gate voltage $V_{G1}$.}
	\label{figsupp}
\end{figure}

{\bf Measurement techniques.} Electrical transport measurements were performed at the base temperature of a dilution refrigerator (electron temperature of 200~mK) using radio-frequency reflectometry \cite{Schoelkopf1998}.  This technique probes the reflection coefficient of a resonant circuit that includes the device as a circuit element (Fig. \ref{figsupp}(a)). As the impedance, in this case the differential conductance, of the device changes so does the reflection coefficient of the resonant circuit (Fig. \ref{figsupp}(b)). This technique allows an increase of bandwidth over a standard DC or lock-in measurement. The sample was embedded in a rf-tank circuit formed by a surface mount 390~nH inductor and a parasitic capacitance (500~fF) to ground.  An rf-carrier signal was applied to the source of the device at the resonant frequency of 360~MHz and the cryo-amplified reflected signal is homodyne detected~\cite{Angus2008}. A bias tee on the sample board permitted the simultaneous measurement of the two terminal DC conductance. The measurements presented here have been obtained after a bias cool down, $V_{G1}$=$V_{G2}$=$V_{TG}$=1~V. When performing transport spectroscopy of the SB device the TG and DB was set well-above threshold ($V_{TG}$=$V_{G2}$=2~V). The devices survived several thermal cycles preserving the subthreshold transitions independently of the bias during the cool-down, confirming its atomic origin.

{\bf Experimental considerations.} Due to the electrostatic configuration of our device in Fig.~\ref{fig1}(b), transport spectroscopy senses only impurities close to the interface immediately below the barrier electrode. The electrostatic properties of single dopants are modified by the presence of an electric field~\cite{Rahman2011a}: For gated donors a reduction in $E_C$ due to screening effects has been experimentally observed~\cite{lansbergen-NatPhys}, while there is experimental evidence of an enhanced $E_C$  in systems under quantum confinement \cite{diarra07}, which is very unlikely to occur for the planar geometry of our device. The possible effect of the dielectric mismatch cannot explain quantitatively in any case the observed energy enhancement~\cite{calderon-Dminus-2010} ruling out any of the aforementioned mechanism as the origin of the simultaneous increase of the binding and charging energies.

In order to obtain highly accurate values of the binding and charging energies we consider the point in $V_{SD}$ at which the edges of the Coulomb diamonds meet. Conventionally a constant voltage-to-energy calibration parameter $\alpha = E_D/V_{G1}$, where $E_D$ is the energy at the dopant site, is used to obtain energy values from gate voltage changes \cite{pierre2009}. However, under the considerations of the constant interaction model, this is expected to be different for different electronic occupations of the molecule as experimentally observed here. Moreover, $\alpha$ could vary with bias conditions \cite{Sellier2006} reducing the accuracy of this method. 

The states a, b in Fig.~\ref{fig1}(c) have been described elsewhere as single dopant states or unintentional quantum dots generated due to potential irregularities at the interface \cite{gonzalez-zalba-NJP2012}.

{\bf Theory.} The theory of H$_2$ molecules in vacuum could be rescaled by the effective semiconductor energy and distance atomic scales, but the results are misleading. In order to get values comparable to the measured ones we have to reach the hypothetical limit of $R\rightarrow 0$ where the two atoms merge into a  He-like atom with $E_C=68$ meV, $E_{\rm I}=56$ meV, and $E_{\rm II}=125$ meV. The equivalent of a He atom in Si would be a group VI donor -- S, Se, or Te -- but the accidental presence of these is highly unlikely and the binding energies for these dopants are much larger~\cite{grimmeiss82} than the ones measured here (e.g. for S, these values are respectively $E_C=295$ meV, $E_I=318$ meV and $E_{II}=613$ meV). Moreover, the singlet-triplet splitting measured here is smaller than the rescaled splitting between ortho and parahelium by two orders of magnitude.

We take the impurity potential $V_{imp}$ to reflect the complications associated with the departure from the hydrogenic model. The deviation of $V_{imp}$ from the point charge screened coulomb potential has its roots on the finite screening length, the structure of the electronic cloud of valence electrons and the covalent bonds. All these effects modify the effective impurity potential at the scale of the atomic radius, which is much smaller than the typical size of the wavefunction of shallow impurities in the hydrogenic model $a^*$. Only the first effect can be taken into account without getting into the details of the chemical nature of the donor. The most common approach is to consider an usual point charge potential modified by a semi-empirical \emph{central cell correction}. This way, we write the full impurity potential $V_{imp}(r)=V_{H}(r)+V_{CC}(r)$ as the superposition of the uniformly screened point charge hydrogenic potential
\begin{equation}
V_{H}(r)=-\frac{e^2}{\varepsilon_{Si} r},
\label{eq:impurity_potential}
\end{equation}
and a central cell correction term $V_{CC}$. Here we take an approach inspired by the Yukawa potential
\begin{equation}
V_{CC}(r)=-\left(1-\frac{\varepsilon_0}{\varepsilon_{Si}}\right)e^{-r/r_{cc}}\frac{e^2}{\varepsilon_0 r}.
\label{eq:central_cell}
\end{equation}
This form gives the correct physical limits for $V_{imp}(r)$ at $r\rightarrow 0$ and $r\rightarrow\infty$, and the characteristic radius of the central cell region $r_{cc}$ is the only empirical parameter in our theory. We take $r_{cc} = 171$~pm to fit the experimental valley splitting. A similar approach was adopted by Wellard and Hollenberg~\cite{wellard05}, with their empirical parameter $\eta$ rescaling the dielectric function to fit the experimental $A_1$ energy.

The multivalley solution to this problem is well known. The lowest energy manifold consists of 1s-like envelopes and combinations of band minima Bloch functions which may be written as
$\Psi^{\mu}=\sum_{i =\pm x, \pm y, \pm z} c ^{\mu}_i \phi_i (\mathbf{r},a_{\mu})$. The particular combinations that transform according to the tetrahedral crystal field ($T_{d}$ symmetry) are $c^{A_1}=(1,1,1,1,1,1)/\sqrt{6}$, $c^{T_x}=(1,-1,0,0,0,0)/\sqrt{2}$, $c^{T_y}=(0,0,1,-1,0,0)/\sqrt{2}$, $c^{T_z}=(0,0,0,0,1,-1)/\sqrt{2}$, $c^{T_x}=(1,1,-1,-1,0,0)/2$ and $c^{T_x}=(1,1,1,1,-2,-2)/\sqrt{12}$. 

These states are split if we take into account the the intervalley coupling due to the crystal field potential, leading to a non-degenerate ground state (A$_1$), a three-fold degenerate first excited state (T) and a doubly degenerate second excitation(E). Nevertheless, the mere inclusion of the intervalley integrals of the Coulomb potential severely underestimates the separation between the excitation energies~\cite{wellard05}.

The most prominent source of splittings comes from the central cell correction discussed above, which acts differently on the six states. While A$_1$ is even under reflections on the cartesian planes around the donor site, T and E are odd, so that in the region where the central cell correction is most active ($r<r_{cc}$), $\Psi^{A_1}$ has a peak while all the $\Psi^{T_i}$ and $\Psi^{E_i}$ have a node, and therefore nearly vanish. Indeed, the experimental binding energies of the $T$ and $E$ states are reasonably described by the hydrogenic model, while the $A_1$  binding energy is considerably larger.

The Kohn-Luttinger model treats this splitting empirically, while no correction to the wavefunction localization is made. Kohn~\cite{kohn} pointed out that the asymptotic behavior of the wavefunction is expected to be significantly altered by the central cell correction, leading to $A_1$ states more localized than the excited states. Following this argument, we ascribe different wavefunction variational decay radii $a_{A1}=a_{CC}=1.20$~nm for the central cell corrected $A_1$ state (obtained variationally from the hamiltonian with the full impurity potential $V_{imp}$ ) and $a_{Ti}=a_{Ei}=a_{sv}=2.14$~nm for all the T and E states (which results directly from solving the hydrogenic problem and disregarding $V_{CC}$).  The variational energy obtained is exactly the experimental energy by construction (by fitting $r_{CC}$). 

The two donors, referred to as $A$ and $B$, located at positions $\mathbf{r_A}$ and $\mathbf{r_B}$ are separated by a vector $\mathbf{R}=\mathbf{r_B}-\mathbf{r_A}$. The crystal defines cartesian directions, so that the energy and wavefunction of an electron at a position $\mathbf{r}$ depends on the full vector and not only on the distances to the point charges $r_A=|\mathbf{r}-\mathbf{r_A}|$ and $r_B=|\mathbf{r}-\mathbf{r_B}|$ --- differently from the analogue H$_2^+$ problem. This is not clear from the effective mass hamiltonian of a double donor, which reads
\begin{equation}
H_{DD} (\mathbf{r})=-\frac{\hbar^2 \nabla^2}{2m^*}+V_{imp}(r_A)+V_{imp}(r_B).
\label{eq:h_1_e_2_d}
\end{equation}
The matrix elements of this hamiltonian are calculated in the molecular orbital approximation, \textit{i.e.}, in the basis set defined by the six valley combinations $\Psi^{\mu}$ centered at $\mathbf{r_A}$ and $\mathbf{r_B}$, so that the hamiltonian is a 12$\times$12 matrix composed of four blocks
\begin{equation}
H=
\left[{\begin{array}{cc}
 H_{AA} & H_{AB}\\
 H_{BA} & H_{BB}
\end{array}}\right]
\label{eq:matrix}
\end{equation}
Blocks $H_{AA}$ and $H_{BB}$ are diagonal and represent the on-site energies, while the tunnel coupling is determined by the blocks $H_{AB}$ and $H_{BA}$. The hamiltonian above is symmetric under the exchange of $A\leftrightarrow B$. Later we will study the system in which this symmetry is broken introducing a detuning potential.

The diagonal blocks read $[H_{AA}]=[H_{BB}]= diag\{ E_{CC}+V^\prime_{imp}, E_{sv}+V^\prime_{H} , E_{sv}+V^\prime_{H}, E_{sv}+V^\prime_{H} , E_{sv}+V^\prime_{H} , E_{sv}+V^\prime_{H}\}$, showing that the onsite energy consists of the single donor binding energies $E_{CC}$ (for A$_1$ states) and $E_{sv}$ (for E and T states), corrected by a long range classical term which reads $V^\prime_{imp}=\langle \Psi^{A_1}(r_A)| (V_H(r_B) + V_{CC}(r_B)|\Psi^{A_1}(r_A)\rangle$ for A$_1$ states and $V^\prime_{H}=\langle \Psi^{A_1}(r_A)| V_{H}(r_B)|\Psi^{A_1}(r_A)\rangle$ for E and T states. It is interesting to notice that, due to the periodicity of the Bloch functions, the same argument regarding the different effect of the central cell on A$_1$ and T/E states apply here. In other words, the $\Psi^{A_1} (r_A)$ state presents a peak at the B site and is influenced by the central cell of the B impurity $V_{CC}(r_B)$, while the $\Psi^{T_i}(r_A)$ and $\Psi^{E_i}(r_A)$ states present a node and are immune to the central cell correction.

The off diagonal terms in $H_{AA}$ and $H_{BB}$ are only approximatly vanishing, since with the presence of the second donor the problem looses its spherical symmetry. But direct calculation reveal this to be a good approximation.

The off-diagonal blocks are related by the hermiticity condition $H^{}_{AB}=H_{BA}^\dagger$. Each term is a summation over integrals of the type
\begin{widetext}
\begin{equation}
\langle \phi_\mu^A|H_{DD}|\phi_\nu^B\rangle = \int  F(r_A, a_i) e^{- i \mathbf{k_\mu} \cdot (\mathbf{r}-\mathbf{R_A})}  u^*_\mu(\mathbf{r}-\mathbf{R_A}) H_{DD} (\mathbf{r}) F(r_B, a_j) e^{ i \mathbf{k_\nu} \cdot (\mathbf{r}-\mathbf{R_B})} u_\nu (\mathbf{r}-\mathbf{R_B}) \mathrm{d}^3 r.
\label{eq:tunnel}
\end{equation}
\end{widetext}

These matrix elements can be straightforwardly calculated using the plane wave expansion of the Bloch functions (see Ref.~\onlinecite{saraiva_intervalley_2011}), and no further approximation is necessary. Nevertheless, it is useful to express these integrals in a simpler form adopting some very robust and well tested approximations: (i) taking the matrix elements for $\mu\ne\nu$ to be vanishing, since these consist of rapidly oscillatory integrands; (ii) taking $u^*_\mu u_\mu \approx 1$, as suggested in Ref.~\onlinecite{saraiva_intervalley_2011}. With these, we write
\begin{equation}
\langle \phi^{A,i}_\mu|H_{DD} (\mathbf{r})|\phi^{B,j}_\nu\rangle = \delta_{\mu,\nu} e^{i \mathbf{k_\mu}\cdot \mathbf{R}} t_{sv}(R,a_i,a_j),
\label{eq:tunnel_simple}
\end{equation}
where $t_{sv}$ is the single valley tunnel coupling
\begin{equation}
t_{sv}(R,a_i,a_j) = \int  F(r_A, a_i) H_{DD} (\mathbf{r}) F(r_B, a_j) \mathrm{d}^3 r.
\label{eq:tunnel_sv}
\end{equation}

The effect of the hopping blocks $H_{AB}$ is analogous to the H$_2$ molecule -- at distances much larger than $a_i$ and $a_j$, this block is null and the states centered around sites A and B are degenerate, while for distances comparable to the wavefunction radius there is the formation of symmetric and antisymmetric superpositions of the localized orbital with splitted energies. Unlike the H$_2$ analogue, it is possible for the antisymmetric state (referred to as antibonding in the context of H$_2$ in vacuum) to be the ground state here, since the hoppings are not necessarily real negative numbers due to the oscillatory phase $exp(i \mathbf{k_\mu}\cdot\mathbf{R})$.

The same oscillatory phase may also lift the degeneracies of the T and E states. For instance, if the pair alignment is along the $x$ direction, the states $T_y$ and $T_z$ are still equivalent, while the state $T_x$ will have a symetric-antisymmetric splitting that oscillates as a function of $R$.

It is known from the H$_2$ problem that the molecular orbital approximation gives accurate results only if the variational wavefunction radius is taken to minimize the expectation value of the complete hamiltonian containing the two protons. We do the same here, minimizing numerically the lowest eigenvalue of the double donor hamiltonian $H_{DD}$ with relation to all the six wavefunction radii $a_i$. 

Given the solution of the problem of one electron with two donors discussed above, we may use these 12 molecular orbitals to write down spin-orbitals from Slater determinants. Combining suitably the Slater determinants, we may identify 78 spin singlets and 66 triplets (for a total of $12 \times 12=144$ two-electron spin-orbitals). These include all possible excitations of the two electrons. Calculating the matrix elements of the full hamiltonian, including the electron-electron repulsion, we may obtain the full configuration interaction ground state. Since we explicitly disconnect the singlet and triplet blocks, this method is often referred to as Spin Adapted Configuration Interaction, or SACI.

The explicit calculation of the two-electron exchange and hybrid integrals is an outstanding problem in computational quantum chemistry, and is often solved approximately. One of the most successful approaches is to fit a number $N$ of gaussian orbitals to the Slater-type orbital obtained in the single electron problem. This basis set, called STO-$N$G, converges quickly~\cite{Hehre1969}. We tested $N=2$ and $N=3$ without significant improvement. All data presented is for $N=3$, though.

Our method is therefore well tested and robust in all regimes except at distances $R$ comparable to the central cell region. At this region, the polarization cloud around the donor nuclei is non-trivial and the effective potential acting on conduction electrons is most likely non-additive. Therefore, this theory is not capable of describing vicinal donors (substituting first nearest neighbor Si sites). We set a minimum distance of $R>1 a_0=0.543$ nm as a conservative boundary for our theory.

\bibliography{ref_vs}

\end{document}